\documentclass[12pt]{iopart}

\usepackage{graphicx}

\begin{document}
\title[Position controlled self-catalyzed growth of GaAs nanowires]{Position controlled self-catalyzed growth of GaAs nanowires by molecular beam epitaxy}

\author{Benedikt Bauer$^1$, Andreas Rudolph$^1$, Marcello Soda$^1$, Anna Fontcuberta i Morral$^2$, Josef Zweck$^1$, Dieter Schuh$^1$ and Elisabeth Reiger$^1$\footnote{elisabeth.reiger@physik.uni-regensburg.de}}
\address{$^1$ Institute for Experimental and Applied Physics, University of
Regensburg, Universit\"atsstr. 31, D-93053 Regensburg, Germany}
\address{$^2$ Laboratoire des Mat\'eriaux Semiconducteurs,
Institut des Mat\'eriaux, \'Ecole polytechnique f\'ed\'erale de
Lausanne, CH-1015 Lausanne, Switzerland}

\begin{abstract} 
GaAs nanowires are grown by molecular beam epitaxy using a self-catalyzed, Ga-assisted growth technique. Position control is achieved by nano-patterning a SiO$_2$ layer with arrays of holes with a hole diameter of 85~nm and a hole pitch varying between 200~nm and 2~$\mu$m. Gallium droplets form preferentially at the etched holes acting as catalyst for the nanowire growth. The nanowires have hexagonal cross-sections with \{110\} side facets and crystallize predominantly in zincblende. The interdistance dependence of the nanowire growth rate indicates a change of the III/V ratio towards As-rich conditions for large hole distances inhibiting NW growth.
\end{abstract}


\submitto{\NT}

\maketitle

\section{Introduction}
Conventional microdevice fabrication techniques are facing more and more restrictions due to resolution limits as the miniaturization of integrated circuits is proceeding. Nanowires (NWs) are discussed as one possible solution to this challenge \cite{Hayden2008}. Prototypes for applications such as NW-based transistors \cite{Bryllert2006} as well as axial or radial heterostructures \cite{Rudolph2009, Caroff2009, Colombo2009, FontcubertaiMorral2008, Tomioka2010} have already been realized. For the next generation of possible devices e.g.\ integrated circuits containing NWs or for photoluminescence experiments on single isolated NWs, control over the position of the NWs is essential. NW growth typically requires the presence of a catalyst, which favors the incorporation of material at the catalyst-NW interface \cite{Wacaser2009, Kolasinski2006}. Position control can be achieved by nano-patterning the catalyzing metal layer with electron beam lithography and metal lift-off method \cite{Maartensson2003}. Using metal-organic vapor phase epitaxy~(MOVPE), NW growth can also be  obtained by selective-area epitaxy, where a patterned SiO$_2$ layers acts as a mask restricting growth to the openings in the SiO$_2$ layer \cite{Noborisaka2005}.

For NW growth in combination with molecular beam epitaxy (MBE) most often gold is used as catalyst \cite{Fan2006, Harmand2005, Wu2002}. More recently a self-catalyzed, Ga-assisted method has been developed with the major advantage that no foreign metal can be incorporated from the catalyst during growth \cite{Colombo2008, FontcubertaiMorral2008b}. Also the crystal structure of the NWs can be tuned, from pure zincblende to 70~\% wurtzite crystal phase, by adjusting the growth parameters \cite{Spirkoska2009}. The gallium droplets which act as catalysts for the NW growth form at nanoscale holes -- small defects -- in an amorphous SiO$_2$ layer. By nano-structuring the SiO$_2$ layer with electron beam lithography (EBL) and wet chemical etching we could turn this randomly distributed self-assembly process into a deterministic growth process where the position of the NWs is pre-defined. Note that the presented technique differs from selective-area epitaxy as here the formation of a catalyst is essential for NW growth. The existing growth model \cite{Colombo2008} is revised, suggesting that also As diffusion on the substrate has to be taken into account to explain the observed NW growth rate dependence.

\section{Experimental details}

For substrate preparation a 15~nm thick amorphous SiO$_2$ layer was sputtered on a (111)B GaAs wafer. Electron beam exposure was carried out in a scanning electron microscope (Zeiss Leo~1530 Gemini extended with a e-beam control unit EK~03 by nanonic) using PMMA as resist. The created etching mask consists of holes with a diameter of approx. 50~nm, which were aligned in a quadratic grid. Fields with five different inter-hole pitches of 200~nm, 250~nm, 500~nm, 1~$\mu$m and 2~$\mu$m were defined. The uncovered SiO$_2$ was etched with ammonium flouride (NH$_4$F) diluted to a concentration of 1.25~\%. Several cleaning steps with acetone, isopropylic alcohol and N-Methyl-2-pyrrolidone (NMP), all with ultrasonic agitation, an oxygen plasma treatment as well as a second NH$_4$F dip were applied to ensure that PMMA residues are completely removed. Due to the isotropic etch process the holes have a final diameter of approximatively (85$\pm$5)~nm.

\begin{figure}
    \includegraphics[width=0.48\textwidth]{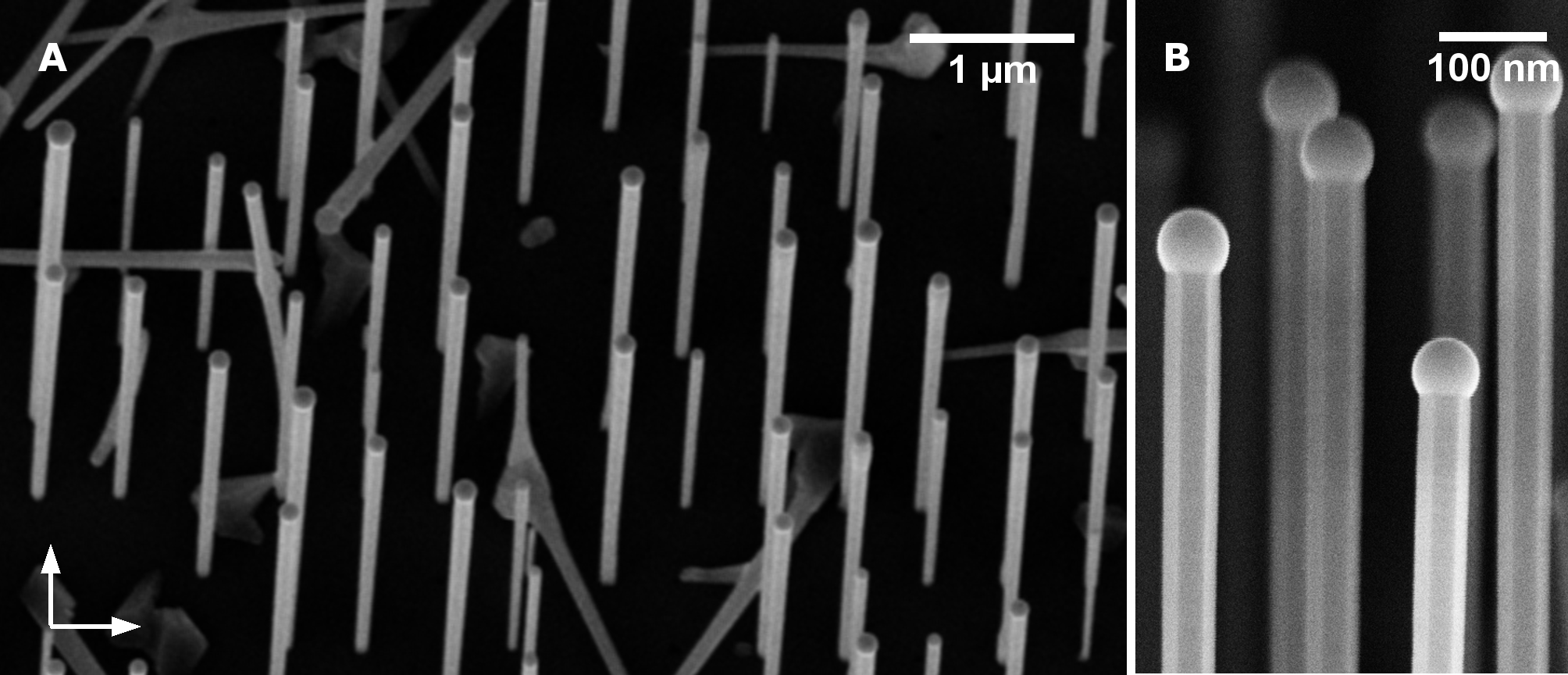}
    \caption{(A) SEM image ($22^\circ$ tilted view) of a NW field     with a relatively high local NW density. The hole pitch of the EBL pattern is 500 nm. The arrows in the bottom left corner indicate the directions of the EBL-defined quadratic grid. (B)~Close-up side view of the NW tips showing the Ga droplets.
    \label{fig:tiltedsem}}
\end{figure}

The NWs were grown using a modified Veeco Gen~II MBE system. The system is equipped with an infrared pyrometer specified for high temperatures. Before growth the substrate is heated to either $580~^\circ$C or $620~^\circ$C for 30 minutes to desorb adhesive molecules from the SiO$_2$ surface as well as to remove oxide from the pre-structured holes. If not other specified following parameter settings were used for NW growth: Substrate temperature: $590-605~^\circ$C, As$_4$ pressure: $5.0 \times 10^{-7}$~Torr, Ga rate: 0.2~\AA/s, yielding an As$_4$/Ga flux ratio of 3.6. The As$_4$/Ga flux ratio can be calculated from the beam equivalent pressure of both beams considering the absolute temperatures, molecular weight, and the ionization efficiency relative to nitrogen of the two specimen \cite{Joyce1993}. An amount of material equivalent to a 300~nm thick 2D~GaAs layer on GaAs(001) was deposited. During growth the substrate was rotated with 7~rpm. After growth both cells were closed immediately and the substrate temperature was reduced below $400~^\circ$C.

\section{Results and discussions}

\begin{figure}
    \includegraphics[width=0.7\textwidth]{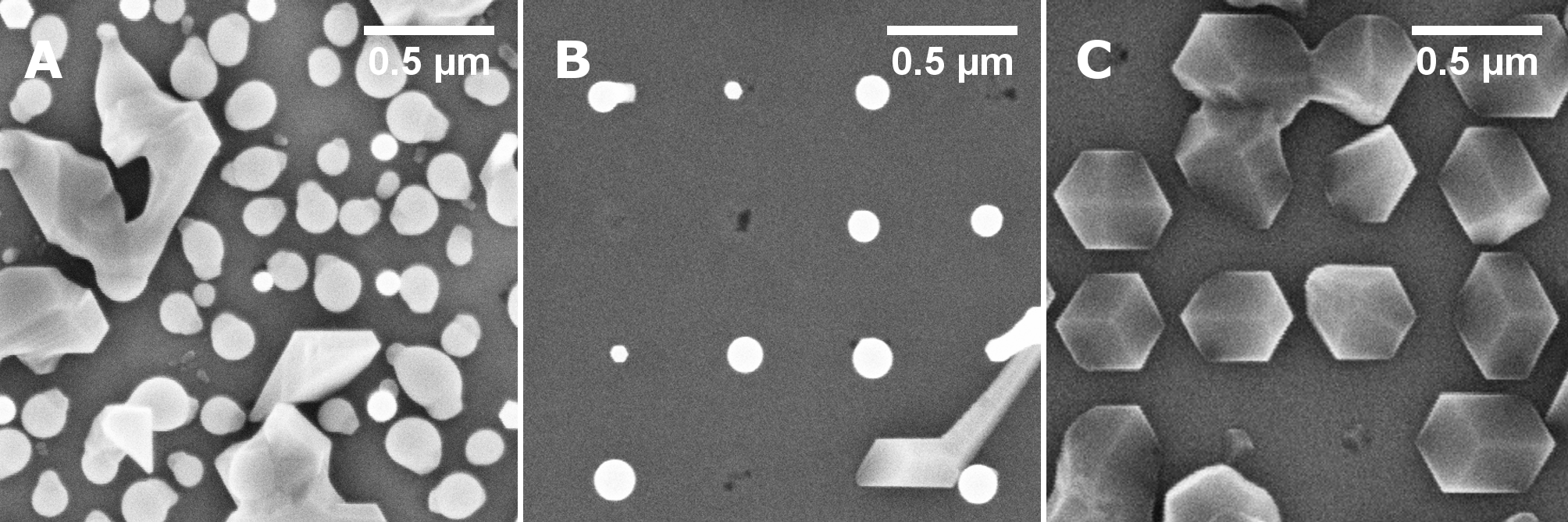}
    \caption{SEM images (top view) of NW fields with 500~nm hole distances grown at different As$_4$/Ga flux ratios. (A)~As$_4$/Ga 2.5: A high density of Ga droplets, but no NW growth is observed. (B)~As$_4$/Ga 3.6: A high NW density is obtained. (C)~As$_4$/Ga 17.9: GaAs crystallites grow.
    \label{fig:asga}}
\end{figure}

The morphology and the crystal structure of the NWs were characterized by scanning electron microscopy (SEM) and transmission electron microscopy (TEM). Figure~\ref{fig:tiltedsem} depicts a SEM image of a NW field with inter-hole distances of 500~nm. The NWs grew preferentially in [111]B direction standing perpendicular on the (111)B oriented GaAs substrate. Occasionally, growth in the $\langle111\rangle$A crystal directions is observed, i.e.\ NWs grow under an angle of $19.5^\circ$ with respect to the substrate (left upper corner of figure~\ref{fig:tiltedsem}(A)). The section of the NW field shown in figure~\ref{fig:tiltedsem}(A) has a rather high local NW density. The NWs are clearly periodically arranged. In the top view SEM image in figure~\ref{fig:asga}(B) and (C) the quadratic grid defined by the EBL pattern can be more easily recognized. NW growth, however, does not occur at all pre-defined positions. The occupation density varies from wafer to wafer -- although processed identically -- but also fluctuates locally within a single pre-structured NW field. This is most likely caused by the non-homogeneity of the wet-chemical etching process. To quantify the success rate of the NW growth process, we estimate the overall NW density to 20~\% taking into account all processed samples. No NW growth is observed outside the pre-defined areas, i.e.\ no growth on unpatterned areas takes place. This is most probable due to the rather thick SiO$_2$ layer. In figure~\ref{fig:tiltedsem}(B) the Ga droplet is clearly visible at the tip of the NWs. The hexagonal shape of the NWs -- also verified by TEM characterization -- is reflected by the three crystal planes of the NW bodies all oriented in $\langle110\rangle$ directions.

We investigated the influence of different pre-growth preparations as well as different growth parameters on the position-controlled growth of the NWs. The etching time of the first etching step as well as the second NH$_4$F dip are crucial parameters. For successful NW growth it is essential that the SiO$_2$ is completely removed down to the GaAs substrate in the etched regions. One reason for the observation that NW growth does not occur at all pre-defined positions is most likely related to the incomplete removal of SiO$_2$, leaving residues inside the holes which do not allow for Ga nucleation. In absence of the second NH$_4$F dip no NW growth could be achieved, instead the formation of GaAs clusters was observed. This indicates that a clean SiO$_2$ layer is essential to ensure that adsorbed atoms diffuse to the etched holes as already seen for Ga assisted growth on unpatterned samples \cite{FontcubertaiMorral2008b}.

Three different As$_4$/Ga flux ratios of 2.5, 3.6 and 17.9 were investigated by varying the As$_4$ beam flux with the Ga rate fixed at 0.2~\AA/s. For an As$_4$/Ga flux ratio of 2.5, the NW density is very low and a huge number of Ga droplets form on the SiO$_2$ layer which do not catalyze NW growth (cf.\ Fig~\ref{fig:asga}). At an As$_4$/Ga flux ratio of 17.9 NW growth is completely suppressed and very regular faceted clusters appear comparable with \cite{Lee2004} which indicates a change of growth regime from self-catalyzed NW growth to selective area epitaxy \cite{Lee2004, Heiss2008, Zhao2009}. An As$_4$/Ga flux ratio of 3.6 leads to NW growth as well as to low cluster formation on the substrate. Note that on the SEM top-view images, as the NWs grow perpendicular to the substrate, only the catalyst droplets of the NWs are visible (white dots in figure~\ref{fig:asga}(B)). Therefore NWs with droplets at their top and simple droplets without NWs are indistinguishable in the top view perspective. By tilting the substrate for taking the SEM pictures NWs with catalyst droplets can be clearly distinguished from simple Ga droplets or clusters.

\begin{figure}
    \includegraphics[width=0.7\textwidth]{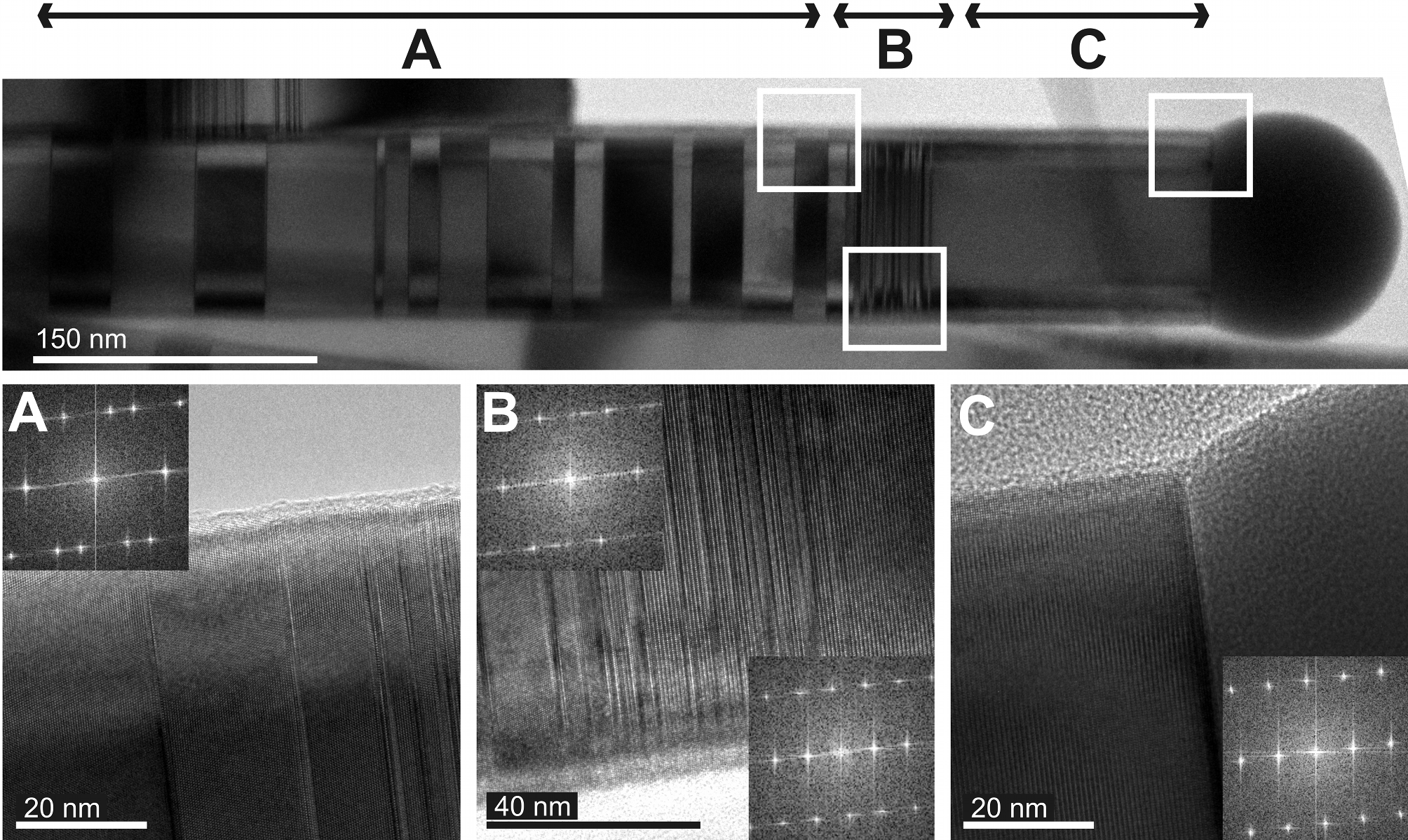}
    \caption{(Top) Bright field TEM picture of a NW oriented in the [110] zone axis. Three regions can be identified. (Bottom) High-resolution TEM images of the different regions and their diffractograms. The main part of the NW~(A) is characterized by the presence of predominantely twinned zincblende segments whereas the part of the NW below the Ga droplet~(C) shows an homogeneous wurtzite crystal structure. Both regions are separated through a transition region~(B) with a high density of stacking faults. \label{fig:tem}}
\end{figure}

The crystal structure of the NWs was investigated by transmission electron microscopy (FEI Tecnai~F30). The NWs were mechanically transferred to a standard TEM grid. All investigated NWs have a hexagonal shape with six \{110\}-oriented side facets. For eight NWs TEM and high resolution TEM (HRTEM) images were taken at various positions along the whole NW length. The dominant crystal structure is zincblende which is in accordance to previous results obtained with this growth technique \cite{Spirkoska2009, Cirlin2010}. 
All NWs consist of three characteristic regions. In figure~\ref{fig:tem} a bright field picture of a typical NW is presented in the upper micrograph where the dimension of the three regions is indicated by arrows. In the bottom line high resolution TEM micrographs and their diffractograms, taken within the marked areas, prove the various crystal structures within the three regions. Region~(C) directly below the Ga catalyst droplet has wurtzite crystal structure. For the shown NW with a diameter of 110 nm this region extends about 150 nm. Transition region~(B), which is located between the wurtzite segment~(C) and the main part of the NW~(A), is characterized by a high density of stacking faults. We attribute regions (B) and (C) to the point when the MBE growth process is terminated. Closing the MBE cells and reducing the substrate temperature does not immediately stop the NW growth, but abruptly changes the growth conditions. This often leads to a change of crystal structure as observed for Au-catalyzed as well as for Ga-assisted NW growth \cite{Glas2007,Cirlin2010}. The main part (A) of the NWs consists predominantly of zincblende segments separated by twin boundaries as shown in figure~\ref{fig:tem}(A). The typical length of this region is several $\mu$m. Since NWs do not necessarily break at their base when mechanically transferred to the TEM grid, an accurate statistic of the length of region (A) cannot be provided. The length of the ZB segments between two twin boundaries varies for different NWs. In particular for NWs with smaller diameters ($<80$~nm) these zincblende segments can extend several 100~nm without any stacking faults present.

\begin{figure}
    \includegraphics[width=0.7\textwidth]{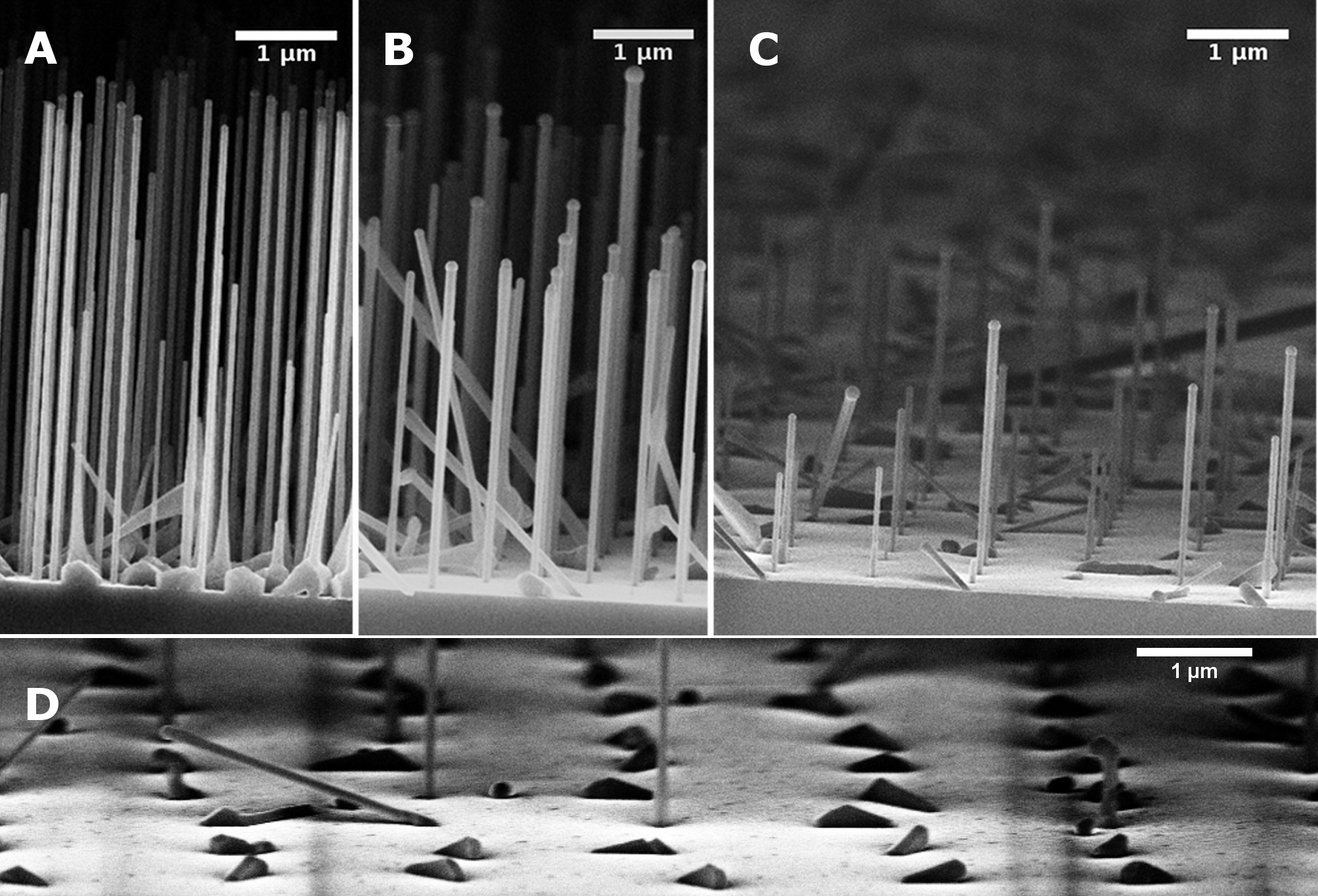}
    \caption{SEM images (side view) of NW fields with different inter-hole pitches of (A)~200~nm, (B)~500~nm (C)~1~$\mu$m and (D)~2~$\mu$m. With increasing inter-hole pitches the NWs growth rate decreases. No NW growth is observed for a hole distance of 2~$\mu$m. \label{fig:sem}}
\end{figure}

In figure~\ref{fig:sem} side view SEM images of NW fields with different inter-hole pitches are shown. NW growth could be achieved for 200~nm up to 1~$\mu$m inter-hole pitches at the pre-defined positions. No NW growth was observed within the 2~$\mu$m field. Although different samples show variations, e.g.\ in NW density, a clear interdistance dependence of the NW growth can be deduced. With increasing hole distances the NW growth rate decreases whereas the  diameter of the NWs increases. For a hole pitch of 200~nm and 250~nm NWs with a diameter between 80~nm to 100~nm and a length of $(5.0 \pm0.5)~\mu$m are observed. This corresponds to a NW growth rate of 1.2~$\mu$m/h. For the 500~nm and 1~$\mu$m NW fields a typical length of $(3.5 \pm0.5)~\mu$m giving a growth rate of 0.83~$\mu$m/h and diameters from 80~nm to 140~nm are observed. The thicker NWs are slightly inverse tapered with a maximum tapering of 1,5~\% whereas NWs with a diameter below 100~nm possess a uniform diameter. In contrast to small hole center distances, where a sharp upper length limit is observable, a higher fluctuation of the NW length and diameter was seen for larger hole center distances.

Interdistance dependence of NW growth has been also reported for NWs grown with metal-organic vapor-phase epitaxy (MOVPE) showing competitive and/or synergetic growth modes \cite{Jensen2004,Borgstrom2007}. In our case, the NW growth rate dependence can be explained by extending the current growth model. Until now, Ga-assisted growth of GaAs NWs was believed to occur as follows: The Ga droplets are continuously filled by Ga atoms adsorbed on the surface that diffuse to the NW sites, whereas the As$_4$ molecules impinge directly on the Ga droplets \cite{Colombo2008}. NW growth only occurs within an appropriate III/V ratio window (see also \ figure~\ref{fig:asga}). This applies not only to the NW growth but also to the Ga droplet formation process (Typically, the formation of Ga droplets takes place under the same growth conditions as the NW growth itself. Attempts to enhance the Ga droplet formation and hence the NW density by opening only the Ga cell before starting the NW growth process did not succeed. No NW growth was observed, instead Ga droplets are formed which do not act as catalyst for NW growth, similar to results obtained at too Ga-rich growth conditions as shown in figure~\ref{fig:asga}(A).).
In this study, we have demonstrated that NW growth within the given growth parameters occurs only provided the inter-NW distance is kept below 1~$\mu$m. For larger distances, GaAs nanoparticles are formed in the holes, but no Ga droplets are observed (cf.\ figure~\ref{fig:sem}) indicating that the III/V ratio has changed to smaller values. In addition, the observed wide variance of NW diameters for NW fields with holes distances larger than 1~$\mu$m is a sign that growth parameters have changed to non-ideal values. This observation can be explained by two additional assumptions to the growth model: (i)~a non-negligible percentage of As$_x$ molecules is adsorbed on the surface and diffuse to the collection/nucleation sites, and (ii)~the collection area of the As$_x$ adsorbates is larger than that of the Ga and both values are in the order of the chosen pitch values. The actual As$_x$/Ga ratio at the catalyst is then not determined by the initial As$_4$/Ga flux ratio but is modified due to the different collection areas for Ga and As$_x$. For hole pitch values below 250~nm we are in the competitive growth regime, where the Ga and the As$_x$ diffusion length exceed the inter-hole distance. In this regime the As$_x$/Ga ratio is constant. When the inter-hole distance lies between the Ga and the As$_x$ diffusion length the As$_x$/Ga value changes towards more As-rich conditions, which leads to the observed decrease of the NW growth rate. This could also induce the observed increase in NW diameter. For hole pitches of 1~$\mu$m to 2~$\mu$m this results in NW growth showing a large fluctuation in lengths up to no NW growth at all. When the hole distance is larger than both diffusion length values the flux ratio will become constant again, however, as the growth conditions are already too As-rich in the intermediate range, this will not lead to NW growth.

In order to check if such a hypothesis is realistic, one can compare the flux of the As$_4$ molecules with the NW growth rate. The highest NW growth rate of 1.2~$\mu$m/h obtained for pitch values of 200~nm and 250~nm corresponds to a flux of $7.3\times10^{18}~$atoms/m$^2$s. This value is factor 16 higher than the Ga flux determined by RHEED oscillations and factor 6 higher than the As$_4$ flux calculated from the beam equivalent pressure. Hence, for Ga-assisted NW growth adsorption and diffusion of both specimen, Ga and As$_4$, has to be considered. Also note that a larger collection area of As with respect to Ga is not in disagreement with the use of Ga-rich conditions. Indeed, the collection area accounts for the maximum distance from which the adsorbates diffuse to the NW. This quantity is not correlated to the desorption rate, which determines the percentage of molecules staying at the surface and contributing to the growth.

\section{Conclusion}
We demonstrated the position-controlled growth of GaAs NWs by nano-patterning a SiO$_2$ layer with e-beam lithography and wet chemical etching. Ga droplets acting as catalyst for the NW growth form at the etched holes. The NWs have a hexagonal form with \{110\} side facets. The dominant crystal structure is zincblende. NW growth was achieved for inter-hole pitches from 200~nm up to 1~$\mu$m. The observed dependence of the NW growth rate on the inter-hole distance is attributed to a change of the As$_x$/Ga ratio at the catalyst droplet. This observation can be explained by adapting the existing growth model. With the nano-patterning approach the benefits of the self-catalyzed growth mode, such as high purity as no metal (catalyst) atoms can be incorporated during NW growth or the tuning of the crystal structure by adapting the growth parameters, can be now combined with position control, which is essential for NW device fabrication. The role of the etched hole diameter on the Ga droplet formation and the NW diameter will be investigated in detail in a separate study. By improving the etching process, e.g.\ using reactive ion etching instead of wet chemical etching, we believe that the NW density can be increased up to 100~\% with NWs growing at all pre-defined sites.

\ack
We thank D.~Weiss for access to the clean room and the chemical lab and W.~Wegscheider for support and discussions. We also thank P.~Atkinson from MPI Stuttgart for the initial cleaning recipe. Our work is partly financed by SFB~689. E.~R. acknowledges financial help from the ERA Nanoscience Project QOptInt. A.~FiM thanks support from the ERC starting grant 'UpCon'.

\section*{References}
\bibliographystyle{unsrt}
\bibliography{PCG}
\end{document}